\documentclass{article}

\usepackage[english]{babel}

\usepackage[letterpaper,top=2cm,bottom=2cm,left=3cm,right=3cm,marginparwidth=1.75cm]{geometry}

\usepackage{amsmath}
\usepackage{graphicx}
\usepackage{authblk}
\usepackage{caption}
\usepackage[colorlinks=true, allcolors=blue]{hyperref}

\usepackage{booktabs}
\usepackage{makecell}
\usepackage[dvipsnames]{xcolor}

\title{Political Biases on X before the 2025 German Federal Election
}
\author[1]{Tabia Tanzin Prama}
\author[2]{Chhandak Bagchi}
\author[2]{Vishal Kalakonnavar}
\author[3]{Paul Krauß}
\author[4,2]{Przemyslaw A. Grabowicz}

\affil[1]{University of Vermont}
\affil[2]{University of Massachusetts Amherst}
\affil[3]{University of Regensburg}
\affil[4]{University College Dublin}

\date{} 

\begin{document}
\maketitle

\setlength{\parskip}{\baselineskip}%
\setlength{\parindent}{0pt}%

\begin{abstract}
This study examines whether German X users would see politically balanced news feeds if they followed comparable leading politicians from each federal parliamentary party of Germany. We address this question using an algorithmic audit tool \cite{piccardi2024reranking} and all publicly available posts published by 436 German politicians on X. We find that the default feed of X showed more content from far-right AfD than from other political parties. We analyze potential factors influencing feed content and the resulting political non-representativeness of X. Our findings suggest that engagement measures and unknown factors related to party affiliation contribute to the overrepresentation of extremes of the German political party spectrum in the default algorithmic feed of X.
\end{abstract}

X users can choose between two content feeds, called For You and Following feeds. The default feed is the For You feed. Both feeds rank content algorithmically, but the Following feed contains posts exclusively of users followed by the feed owner, prioritizing recent content. By contrast, only 20\% of posts shown in the For You feed are from the users they are following.

We created two likewise sock-puppet X accounts,\footnote{There was no difference between the two sock puppet accounts other than their names and associated gender. Their feeds were nearly the same. Our analysis combines information from the For You feeds of the two accounts by averaging the numbers of post appearances between the two accounts.} made them follow the same set of German politicians, and tracked their feeds for 4 weeks, from January 3 noon to January 31 noon German time, by opening and saving the first page (35 posts) of each of their feeds every half an hour. The sock puppets followed 64 users on X – seven to nine members from each of the eight major German federal parliamentary parties (ordered from left-wing to right-wing: BSW, Linke, SPD, Greens, FDP, CDU, CSU, AfD). To better understand the factors that determine post appearance in the For You feed, we collected tweets of all German parliamentarians who have X accounts (436 politicians) and other users most appearing in the For You feed (69 users), including the 64 politicians followed by the sock-puppet accounts (see an overview in Table \ref{tab:german_party_tweets}).

\begin{figure}[ht]
    \centering
    \includegraphics[width=0.8\textwidth]{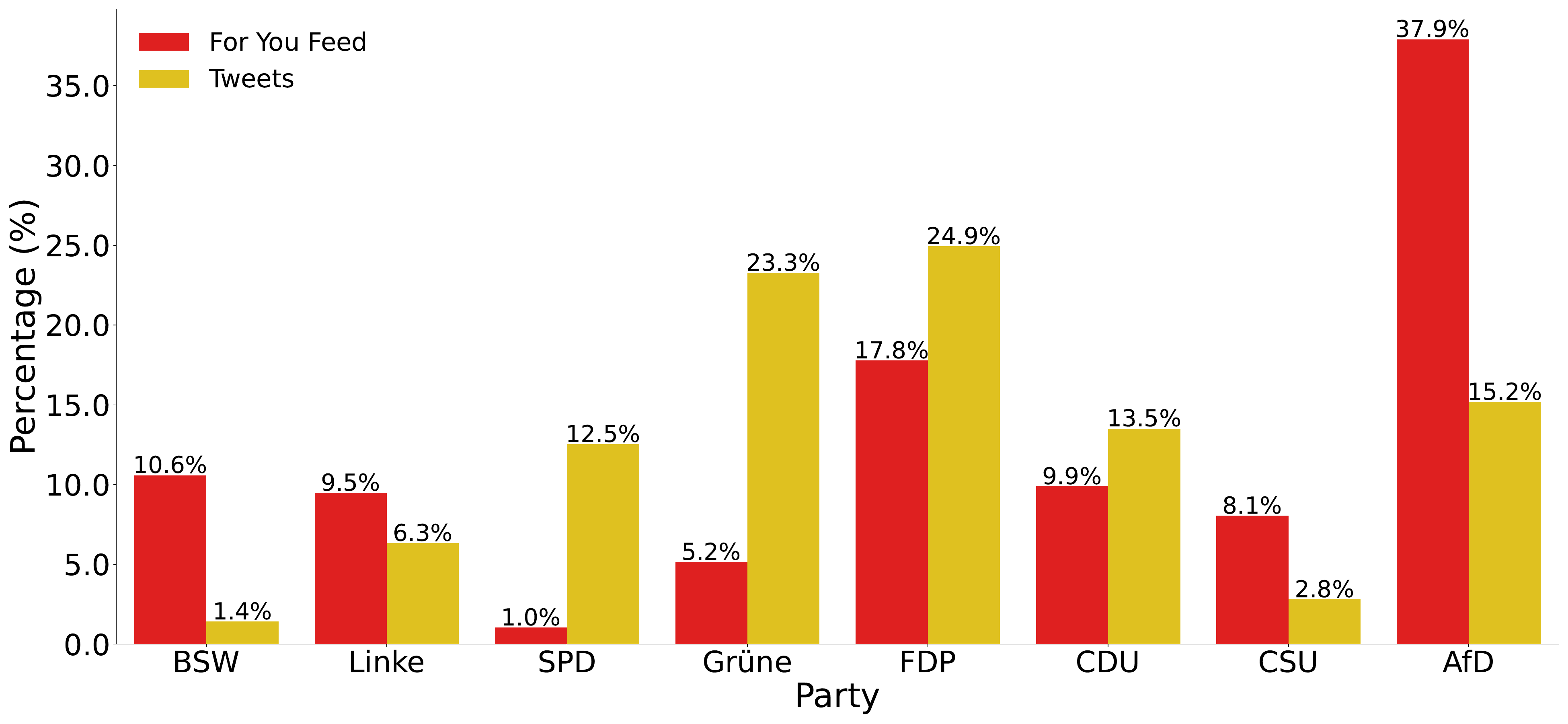}  
    \caption{The fraction of the For You feed occurrences (red bars) and posts on X (yellow bars) in January 2024 of the 436 members of eight German political parties, ordered by their political ideology: from far left (BSW) to far right (AfD).}
    \label{fig:for_you_feed}
\end{figure}

We find that the posts of the AfD politicians appeared most frequently in the For You feed, accounting for about 37.9\% of posts in the feed, even though the AfD politicians posted only 15.2\% of the tweets by politicians during this time (see Figure \ref{fig:for_you_feed}). Similarly, the posts of the other German populist party, the left-wing BSW, appeared 10.6\% times in the For You feed, despite the fact that their politicians created only 1.4\% of politician tweets. 

These results show that X disproportionately highlights extremes of the German political spectrum, in particular the populist parties: left-wing BSW and right-wing AfD. For the three largest non-populist German parties – CDU, SPD, and Greens – the results are very different. Their members appeared less frequently in the For You feed in January than they tweeted. SPD politicians, who are leading the current German government, appeared in the For You feed 12 times less frequently than they posted: 1\% vs. 12.5\%. Representatives of CDU (leading  in opinion polls), appeared in the For You feed 9.9\% of the time, even though they created 13.5\% of tweets, and Greens appeared 5.2\% of time in the feed, while creating 23.3\% of politician tweets.

Because most of the posts in the For You feed are from users other than politicians, we analyzed who appeared most frequently in this feed. The top four most-appearing users are: (1) Elon Musk, who supports AfD, (2) Alex Jones, an American far-right radio show host, and (3-4) Dennis Hohloch and Stephan Brandner from AfD. Among post appearances of the top 100 most appearing users, 48\% are from users who are either AfD members or supporters of AfD, far-right ideology, or conspiracy theories (see the classification in \href{https://docs.google.com/spreadsheets/d/19Uk7SViGdQyo-SThdxM13F-kkqR4qaQF6-ffQfb8uE4/edit?usp=sharing}{Supplementary Information}).

\begin{figure}[!b]
    \centering
    \includegraphics[width=0.8\textwidth]{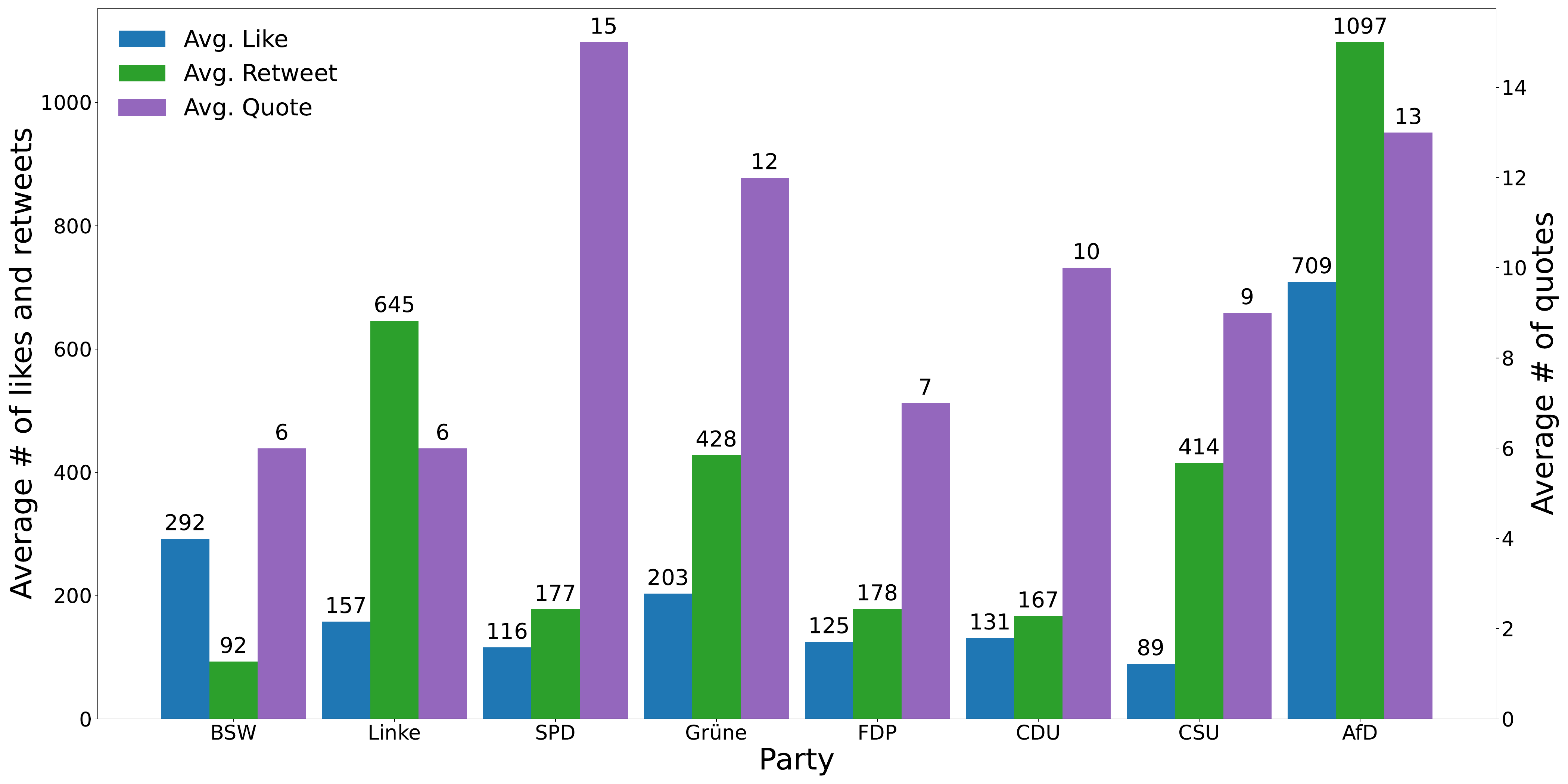}  
    \caption{The number of likes, retweets, and quote tweets per average post published in January 2024 by the 436 German parliamentarians grouped by their political party.}
    \label{fig:Engagement}
\end{figure}

These differences in feed appearances are due to complex reasons. The For You feed algorithm promotes more engaging posts and the members of AfD and BSW may receive more likes and retweets than the other parties, while SPD, CDU, and Greens may draw less engagements. Indeed, an average post of AfD members tends to receive more likes and retweets, as shown in Figure \ref{fig:Engagement}. However, the posts of SPD, CDU, and Greens tend to receive similar numbers of engagements as other non-AfD parties.

\begin{figure}[t]
    \centering
    \includegraphics[width=0.9\textwidth]{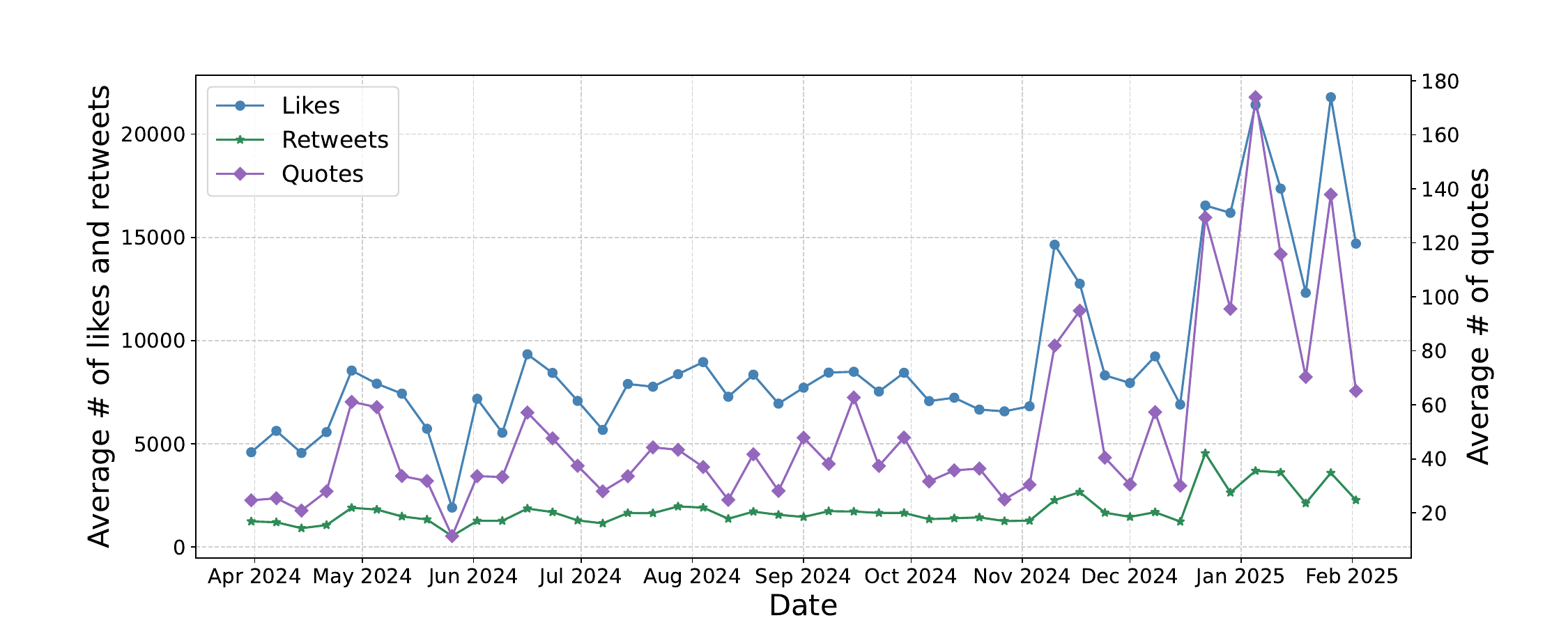}  
    \caption{The number of likes, retweets, and quote tweets per average post of AfD leader, Alice Weidel, published between March 25, 2024, and February 1, 2025.}
    \label{fig:alice_weidel}
\end{figure}

\begin{figure}[!b]
    \centering
    \includegraphics[width=0.9\textwidth]{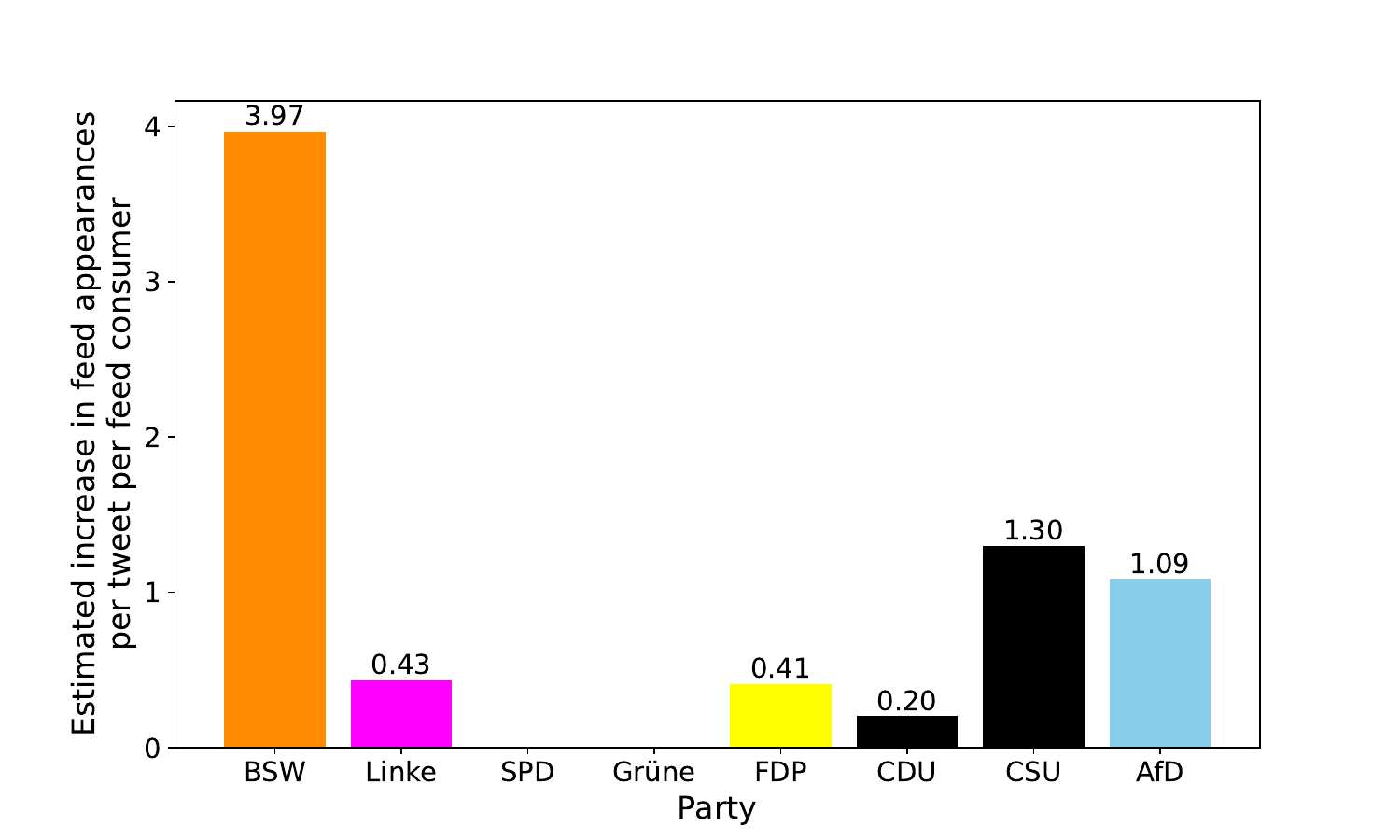}  
    \caption{The estimated significant increases, per tweet per feed consumer, in the For You feed appearances for each of the German parties, after taking into account differences in the numbers of engagements and engagement ratios. We estimate these increases with respective regression coefficients. For the SPD and Greens they are not significantly different from zero. All other estimated increases are significant (p-values below 0.001). The differences in the numbers of tweets per party do not affect these results, since the estimates are computed per tweet.}
    \label{fig:regression_analysis}
\end{figure}

Furthermore, the number of engagements with AfD tweets between mid-November and mid-December was about half of that in January. On the 20th of December Elon Musk started to publicly support AfD, writing on X that “only the AfD can save Germany” and promoting its leader, Alice Weidel, which resulted in a spike in the number of engagements with Weidel’s posts (Figure \ref{fig:alice_weidel}). Such engagements may also be generated by bots and overrepresent certain groups of users, e.g., bots and young right-leaning individuals were more likely to engage with the U.S. presidential election content on X~\cite{scarano2024support, scarano2025election}.

Finally, prior research suggests that certain types of social media posts are more likely to attract views and reactions, namely posts containing misleading information and conspiracy theories, because they often sound sensational and novel \cite{vosoughi2018spread}. If platforms do not inform users about such content in a timely manner, then political parties can increase their visibility by posting more sensational and misleading content.

\begin{table}[!t]
    \centering

    \begin{tabular}{lccc}
        \toprule
        \textbf{German} & \textbf{} & \textbf{} & \textbf{Accounts} \\
        \textbf{party} & \textbf{Accounts} & \textbf{Tweets} & \textbf{ followed} \\
        \midrule
        AfD & 67  & 5013  & 8 \\
        BSW & 13  & 489   & 7 \\
        CDU & 72  & 3281  & 9 \\
        CSU & 22  & 771   & 7 \\
        FDP & 86  & 6475  & 8 \\
        Grüne & 89  & 4884  & 8 \\
        Linke & 27  & 1654  & 8 \\
        SPD & 71  & 2808  & 9 \\
        None & 58  & 21322 & 0 \\
        \bottomrule
    \end{tabular}
    \caption{The number of analyzed users per German political party they represent or endorsed, and the number of tweets they created between January 3 and January 31, 2025. Overall, we analyzed 505 users, including 436 German parliamentarians and 69 other users who were among the top 100 most appearing in the For You feed, 11 of whom represented a party account or endorsed a party. Due to data access limits, we were not able to complete post collection for 23 out of the 505 users (listed in \href{https://docs.google.com/spreadsheets/d/19Uk7SViGdQyo-SThdxM13F-kkqR4qaQF6-ffQfb8uE4/edit?usp=sharing}{Supplementary Information}). 
}
    \label{tab:german_party_tweets}

\end{table}

Even if we account for the differences in the average number of engagements per post of each party member, the For You feed algorithm may, directly or indirectly, advantage the populist parties. To investigate this possibility, we regressed the number of For You feed appearances of each tweet published by the studied users this January against the following factors: counts of views, likes, retweets, quotes, replies, as well as their ratios with the number of views, whether the user is followed, and which political party they belong to (see Model 1 in Table \ref{tab:regression_results} below). The model shows that the counts of likes and quotes are the most important factors positively related to the appearance in the For You feed. However, the model also attributes a significant positive relationship between political party affiliation and the feed appearance. The posts of BSW, CSU, and AfD affiliates appeared in the feed significantly more often (p $<$ 0.001) than one would expect based on their engagement counts and ratios and the other regression factors. SPD and Greens did not enjoy such significantly higher than expected feed appearances (see significant regression coefficients in Figure \ref{fig:regression_analysis}). 

The results of this regression model suggest that party affiliation is related to the appearance in the For You feed through other factors than the engagement measures. These factors contribute to the overrepresentation of extremes of the German political party spectrum in the For You feed, but we are unable to further characterize them using public information from X.  

Also, this regression model has important limitations. First, it explains only about 7\% (R2 = 0.07) of the variance in the feed occurrence counts (Model 1 in Table \ref{tab:regression_results} below). The remaining 93\% of the variance corresponds to the factors that are not included in the model, such as: text and images of posts, interests and characteristics of feed owner, trending topics, and the likelihood of the feed owner to engage with given posts. A more accurate and realistic model would require access to the code of the For You feed algorithm and its input data. However, the last time that \href{https://github.com/twitter/the-algorithm-ml/blob/main/projects/home/recap/README.md}{code} was released was in 2023 and the release did not include any information about its input data. Such data is not public and it is practically impossible to access it without cooperation with X. Furthermore, according to reports, the algorithm was significantly \href{https://www.theverge.com/2023/2/14/23600358/elon-musk-tweets-algorithm-changes-twitter}{altered} in 2023 and July 2024~\cite{graham2024computational}, promoting Elon Musk’s content, so it is not clear how it works now. Second, the same regression model–except without party affiliation–explains about 6\% (R2 = 0.06), rather than 7\%, of the variance in the feed occurrence counts (Model 2 in Table \ref{tab:regression_results} below). This implies that unknown factors related to party affiliation explain about 1\% of the variance in the feed occurrence counts across all parties. While this percentage is small, it corresponds to a statistically significant increase in appearances of tweets in the For You feed of a sock puppet user (or a feed consumer resembling them). For BSW, the increase corresponds to 3.97 appearances more per tweet per feed consumer; for AfD, it is 1.09 appearances more per tweet per feed consumer (Figure \ref{fig:regression_analysis}). These numbers can become large once we multiply them by the number of tweets produced by the representatives of these parties (about 5,000 tweets per month for AfD) and the number of feed consumers (possibly all active German users of X interested in politics).

\begin{table}[!t]
    \centering

    \begin{tabular}{l | rr | rr}
        \toprule
        \textbf{Factor} & \multicolumn{2}{c|}{\textbf{Model 1}} & \multicolumn{2}{c}{\textbf{Model 2}} \\
        \midrule
        Party = AfD            & 1.09  &$^{ ***}$  & -  &  \\
        Party = BSW            & \textbf{3.97}  &$^{***}$  & -  &  \\
        Party = CDU            & 0.20  &$^{*}$  & -  &  \\
        Party = CSU            & 1.30  &$^{***}$  & -  &  \\
        Party = FDP            & 0.41  &$^{***}$  & -  &  \\
        Party = The Greens     & -0.04         &  & -  &  \\
        Party = The Left       & 0.43  &$^{***}$  & -  &  \\
        Party = No Party Affiliation & 0.19 &$^{***}$  & -  &  \\
        Party = SPD            & -0.13         &  & -  &  \\
        Favorite Count         & \textbf{37.81} &$^{***}$  & \textbf{36.57} & $^{***}$   \\
        Retweet Count          & -0.45         &  & 0.28  &  \\
        Reply Count            & \textbf{2.79}          &  & \textbf{2.07}  &  \\
        Quote Count            & \textbf{14.59} &$^{***}$  & \textbf{14.74} &  $^{***}$  \\
        Views Count            & \textbf{4.55}   &$^{**}$  & \textbf{4.94}  & $^{**}$   \\
        Favorite Ratio         & 0.018 &  $^{*}$    & 0.043 & $^{***}$  \\
        Retweet Ratio          & $-7.4 \times 10^{-9}$  &  & $-1.1 \times 10^{-8}$  &  \\
        Reply Ratio            & 0.015  & $^{***}$  & 0.016  &$^{***}$  \\
        Quote Ratio            & 0.071 &  $^{***}$  & 0.070  & $^{***}$  \\
        Followed by User       & 1.68 & $^{***}$    & \textbf{2.12} & $^{***}$    \\
        \midrule
        Number of Observations (\(N\)) & 46,697 &  & 46,697 &  \\
        \( R^2 \) & 0.07 &  & 0.06 &  \\
        Adjusted \( R^2 \) & 0.07 &  & 0.06 &  \\
        \bottomrule
    \end{tabular}
   \caption{Regression of the number of For You feed appearances of each tweet published by the studied users in January. Model 1 analyzes the impact of counts of views, likes, retweets, quotes, replies, their ratios with the number of views, whether the user is followed, and the political party affiliation of the user. Model 2, in contrast, examines the same factors but excludes political party affiliation. Both models consider all 46,697 tweets from 436 German politicians and the top 100 most frequently appearing users in the For You feed. As the target variable of the regression, we use an average over the two sock puppet accounts of the number of appearances in their For You feeds. To compare the importance of different factors, we rescaled all of the factors so that they take values between 0 and 1. The top 5 most important factors are in bold font. We indicate statistical significance at levels \( p<0.001 \) (\( ^{***} \)), \( p<0.01 \) (\( ^{**} \)), and \( p<0.05 \) (\( ^{*} \)).}
      \label{tab:regression_results}
    
\end{table}

Overall, such disparities in the exposure and engagement with content of different parties can translate to differences in support for political parties. We identified more than fifty X polls estimating support at the federal level for German political parties since the collapse of the German government coalition on November 6, 2024. The results of these polls show largely more support for AfD than for its key competitor, CDU, which leads in traditional opinion polls by about 8\% points over AfD. On average, X polls show 44\% support for AfD, 38\% for Greens, 12\% for CDU, and 6\% for SPD. Such polls are not representative and generally reflect biases in engagements, as our prior peer-reviewed research shows for thousands of U.S. election polls on X \cite{scarano2024support, scarano2025election}. Most importantly, biased social media content, if presented as representative, can be deceiving and can be used to influence public discourse and opinion in an undesired way.

To complete our understanding of such striking differences in exposure and engagement with political content, and maintain our democratic societies informed about biases in social media platforms, there is a need for further social media research. The European Union's Digital Services Act (DSA) of 2024, gives vetted researchers a way to access data to study systemic risks to society from online platforms, but it is not clear whether platforms will fully comply. On the one hand, the U.S. tech industry and Trump's second administration seem to \href{https://www.nytimes.com/2019/10/07/business/tech-shield-trade-deals.html}{oppose} such regulation, while arguing against censorship. On the other hand, Trump's first administration \href{https://www.cnn.com/2019/08/09/tech/white-house-social-media-executive-order-fcc-ftc/index.html}{drafted} an executive order requiring online platforms to certify their political neutrality. Platform transparency and accountability regulation, such as DSA, if fully implemented, would allow researchers to study and characterize representativeness and impartiality of online platforms.

\section*{Acknowledgements} We thank Maria Grabe, JungHwan Yang, Gunnar Krüger, Nathan Niedermeier, and Markus Reichert for their valuable comments and feedback.

\bibliographystyle{ieeetr}
\bibliography{main}

\end{document}